\input{aipcheck}

\documentclass[
    ,final            
  ]
  {aipproc}

\layoutstyle{6x9}

\def\msun{{\rm M_{\odot}}}

\def\be{\begin{equation}}
\def\ee{\end{equation}}

\def\msun{{\rm M_{\odot}}}

\begin{document}

\title{Magnetic Fields, Accretion, and the Central Engine of Gamma--Ray Bursts}


\author{A.R. King}{
  address={Theoretical Astrophysics Group, University of Leicester,
Leicester LE1 7RH}
}

\begin{abstract}

I briefly review magnetic effects in accretion physics, and then
go on to discuss a possible central engine for gamma--ray bursts.
A rotating black hole immersed in a non--axisymmetric magnetic
field experiences a torque trying to align spin and field. I
suggest that gamma--ray burst hosts may provide conditions where
this effect allows rapid extraction of a significant fraction of
the hole's rotational energy. I argue that much of the
electromagnetic emission is in two narrow beams parallel and
antiparallel to the asymptotic field direction. This picture
suggests that only a mass $\sim 10^{-5}\msun$ is expelled in a
relativistic outflow, as required by the fireball picture.

\end{abstract}

\maketitle


\section{Accretion}

Accretion -- the conversion of gravitational potential energy to other
forms -- is the most efficient energy source in nature. Accordingly it
is a prime candidate for powering the most luminous systems: at the
galactic scale these are X--ray binaries and ultraluminous X--ray
sources (ULXs), while the most powerful distant sources are quasars
and active galactic nuclei (AGN). Dropping a mass $\Delta M$ on to a
compact oject of mass $M$ and radius $R$ releases energy
\begin{equation}
\Delta E = {GM\Delta m\over R}
\end{equation}
where $G$ is the gravitational constant. Thus the accretion yield
increases with {\it compactness} $M/R$. But matter does not in
general want to fall into a very small radius $R$, just as the
Earth does not fall into the Sun. Just like the Earth, matter
which might otherwise accrete tends to orbit at a radius given by
its angular momentum The major problem in accretion theory is
getting matter to lose enough angular momentum to accrete. In an
active galaxy, the matter supply probably has specific angular
momentum of order $j \simeq (GMa)^{1/2}$, where $M$ is the
supermassive black hole mass and $a$ is a length order parsecs or
more. To get this matter to accrete to the black hole one must
reduce $j$ to a value $j_{\rm acc} \simeq (GMr)^{1/2}$ where $r$
is of order the hole's Schwarzschild radius $\sim 10^{14}$~cm.
Thus $j$ has to be reduced by a factor at least $(a/r)^{1/2}
\simeq 0.003$. Similar considerations hold in close binaries.

The agency which transports angular momentum outwards to allow
matter to moves inwards is usually (and perhaps misleadingly)
called {\it viscosity}. Until recently there was little progress
in identifying a suitable mechanism. The main problem is that
purely hydrodynamical processes are generally sensitive to the
angular momentum itself, and thus transport angular momentum from
large values to small, i.e. inwards. Instead we need a process
which is sensitive to the matter {\it velocity}, which does
increase inwards, and so would lead to angular momentum transport
outwards. Magnetic fields naturally suggest themselves, as they
offer a way of directly connecting fast--moving (low a.m.) matter
at small radii to slow--moving (high a.m.) matter at larger radii.
As shown by Balbus and Hawley (1991), the magnetorotational
instability (MRI) discovered originally by Velikhov (1959) and
Chandrasekhar (1961) offers a mechanism for this process. The
physics of the process is -- with hindsight -- straightforward. A
fieldline connecting fast--rotating matter close in to the
accretor with slower matter further out is dragged in such a way
that it rotates more slowly (rapidly) than its surroundings at
small (large) radii, and so tends to move further in (out).
Eventually differential rotation creates an azimuthal field which
is now unstable to the Parker instability -- the gas pressure and
density inside its flux tubes are lower than outside, so they are
buoyant and tend to rise, recreating a vertical field so that the
cycle starts again.

We thus have a picture of disc accretion in which tangled chaotic
magnetic fields transport angular momentum outwards and drive
matter in. In general the field directions of neighboring disc
annuli are uncorrelated, so now large--scale organised field
arises. However if we wait long enough it must happen by change
that all the field directions in a patch of disc happen to line
up. The effect of this was recently studied by King et al. (2004).
Such an organised field is more effective in transporting angular
momentum, so more matter moves in. But this motion drags in the
field lines and tends to strengthen the field. Eventually this
process amplifies the field to the point where instead of
diffusing inwards, matter moves inwards as a wave (formally the
diffusion equation for mass inflow changes into a wave equation).
This wave of matter is accompanied by a strong magnetic field, and
moves in a relatively non--dissipative manner. It is tempting to
see these conditions as suitable for launching a jet. This may be
how accretion discs are able to produce jets, and indeed sometimes
alternate the processes of accretion and jet production, as in the
microquasar GRS~1915+105 (Belloni et al., 1997).

\section{Accretion in Magnetic Binary Systems}

In the last Section I described how magnetic fields may drive
accretion. If the compact accretor itself possesses an intrinsic
magnetic field this itself can act as an obstacle to the flow.
This occurs when  the typical radius $R_{\rm mag}$ at which the
field is still dynamically important is comparable with various
lengthscales within the binary, all related to the separation $a$.
As a measure of $R_{\rm mag}$ we can take the Alfv\'en radius
given by equating magnetic stresses $\mu^2/8\pi r^6$ with material
ones $\rho v^2$, where $\mu = BR^3$ is the magnetic moment of the
accretor.

The no--hair theorems forbid intrinsic fields for black holes, so
these ideas are only applicable in binaries containing neutron
stars and white dwarfs. It is important to realise that the latter
have stronger magnetic moments $\mu = BR^3$ than than the former:
for white dwarfs we can have $B$ as large as $10^7 - 10^9$~G and
hence $\mu \sim 10^{34} - 10^{36}$~Gcm$^3$ (although the latter
are rare), whereas a neutron star with $B = 10^{12}$~G will only
have $\mu \sim 10^{30}$~Gcm$^3$.

The relative importance of the magnetic moment in a binary is measured
by the quantity $\mu/a^3$, where $a$ is the binary separation. Ranking
systems by this quantity defines a hierarchy of accretion
flows. Observationally they are distinguished by the ratio of spin to
orbital periods.

\subsection{Small $\mu/a^3$}

For systems with small $\mu/a^3$ (essentially all neutron star
binaries, binary white dwarfs with weak fields, or wide WD systems
with strong $\mu$) disc formation is unaffected by the field,
because $R_{\rm mag} << R_{\rm circ}$, where $R_{\rm circ}$ is the
circularization radius -- the Kepler orbit with the same specific
a.m. as the matter transferred from the donor star. In this
situation the accretor gains only low specific a.m. $\sim
(GMR_{\rm circ})^{1/2}$ and spins up to short spin periods $P_{\rm
spin} \sim 10 - 100$~s where $R_{\rm mag} = R_{\rm co}$, the
corotation radius, where the field lines rotate at the local
Kepler value.

This is the typical situation in high--mass X--ray binaries,
and also occurs in just two cataclysmic variables: the wide CV GK Per
(orbital period 2 days) and the very weak--field system DQ~Her. There
is one exceptional CV: AE Aquarii has $P_{\rm spin} = 33$~s, but no
accretion disc. The accretion rate has dropped sharply in the recent
past (as the system's secondary/primary mass ratio decreased below
$\sim 1$) and the rapid WD spin can now expel the transferred matter
via propeller action.

\subsection{Modest $\mu/a^3$}

When $\mu/a^3$ is large enough that $R_{\rm mag} > R_{\rm min} =
0.5R_{\rm circ}$ the accretion stream from the donor hits the
accretor's magnetic field before it can orbit and make an accretion
disc. In this case the WD spins up until $R_{\rm co} \simeq R_{\rm circ}$.
This implies a relation between spin and orbital periods of the form
$P_{\rm spin} \simeq 0.08P_{\rm orb}$, the precise coefficient
depending on the binary mass ratio. This situation is possible only
for white dwarfs, and these systems are called intermediate polars.

\subsection{Slightly higher $\mu/a^3$}

For slightly larger $\mu/a^3$, the field influences the matter
directly issuing through the inner Lagrange point $L_1$. The white
dwarf spins up until the corotation radius is equal to the distance to
$L_1$. This leads to a range of equilibria with $P_{\rm spin} \sim
0.6P_{\rm orb}$ or higher. These are the EX Hydrae systems.

\subsection{High $\mu/a^3$}

At the highest values of $\mu/a^3$, the field interacts directly
with the donor, and forces the whole dwarf to corotate (or nearly
so) with the binary orbit, i.e. $P_{\rm spin} \simeq P_{\rm orb}$.
These are the AM Herculis systems.

\section{The Central Engine of Gamma--Ray Bursts}

Gamma--ray bursts liberate a significant fraction of the rest--mass
energy of a star (i.e. $E_{\rm burst} > 10^{51}$~erg~s$^{-1}$) over
intervals ranging from a few seconds to minutes.  The fireball picture
(Rees \& Meszaros 1992) explains the otherwise puzzling ability of
such sources to vary on short timescales by arguing that the energy of
the burst drives a relativistic outflow with a bulk Lorentz factor
$\gamma \sim 100$. Rest--frame variations of the
central engine on timescale $t_{\rm var}$ are then seen in the lab frame
to have timescales
\begin{equation}
t_{\rm lab} \simeq {1\over 2\gamma^2}t_{\rm var}
\label{var}
\end{equation}
To produce  such behaviour the baryonic mass $M_{\rm out}$ of the
outflow must obey $E_{\rm burst} \sim \gamma M_{\rm out}c^2$, so
that
\begin{equation}
M_{\rm out} \sim 10^{-5}\msun
\label{1}
\end{equation}
This `baryon--loading' constraint is quite stringent, as many
models of the central engine suggest instead that the prompt
energy release may energise a larger mass. The most promising way
of satisfying (\ref{1}) appears to involve prompt energy release
as pure Poynting flux, such as may result from the
Blandford--Znajek (BZ) process (Blandford \& Znajek, 1977).

Other mechanisms are possible. The torque between a spinning black
hole and a nonaxisymmetric magnetic field (King \& Lasota, 1977)
releases a large fraction of the rotational and thus rest--mass energy
of the hole, and was recently recently reconsidered by Kim et
al. (2003). Here I study this process further. I argue that
electromagnetic energy $E_{\rm burst}\sim 10^{51}$~erg~s$^{-1}$ is
released largely in two narrow oppositely--directed jets, with
characteristic diameter of order the ergosphere radius. This
arrangement satisfies the constraint (\ref{1}).

\section{The Alignment Torque}

Press (1972) deduced the existence of a torque on a spinning black
hole immersed in a non--axisymmetric magnetic field as a corollary of
Hawking's (1972) theorem that a stationary black hole is either static
or axisymmetric. Press was able to calculate the torque for the
simpler case of a scalar field uniform at infinity, assuming that the
hole and field passed through a sequence of stationary states, and
inferred an answer for the magnetic case. King and Lasota (1977)
calculated the magnetic torque explicitly, with the result
\begin{equation}
{\bf N} = {2G^2\over 3c^5}M({\bf J}\wedge{\bf B})\wedge{\bf B}
\label{2}
\end{equation}
(where $M$ is the hole mass, ${\bf B}$ the magnetic field at
infinity and ${\bf J}$ the hole's angular momentum), a factor 2 larger
than Press's estimate.

The expression (\ref{2}) shows that a black hole aligns its spin
with a stationary magnetic field by suppressing the angular
momentum component $J_{\perp}$ exponentially on a timescale $\tau
\simeq J/N$, i.e.
\begin{equation}
J_{\perp} = J_{\perp,~0}e^{-t/\tau}
\label{3}
\end{equation}
 with
\begin{equation}
\tau = {3c^5\over 2G^2MB^2}
\label{4}
\end{equation}
The form of (\ref{2}) means that there is no precession as this
occurs. The parallel component $J_{\|}$ remains fixed, so that the
total angular momentum $J = |{\bf J}|$ decreases on the timescale
(\ref{4}). This process extracts rotational energy $E$ from the hole,
since
\begin{equation}
\dot E \propto {\bf J}\cdot{\bf N} \propto ({\bf J}\wedge
               {\bf B})\wedge {\bf B}\cdot {\bf J}
 = -[J^2B^2 - ({\bf J}\cdot{\bf B})^2] = -J_{\perp}^2B^2.
\label{en}
\end{equation}
In agreement with these ideas we note that since
\begin{equation}
J = M{Ga\over c}  = Ma_*{GM\over c} = Ma_*cR_g
\label{j}
\end{equation}
where $a_* = a/M$ is the dimensionless Kerr parameter and $R_g =
GM/c^2$ the gravitational radius, we can write
\begin{equation}
N \sim R_g^3B^2 \sim R_g{L\over c},\ \
\tau \sim {J\over N} \sim {a_*Mc^2\over L}
\label{ntau}
\end{equation}
where
\begin{equation}
L \sim R_g^2B^2c
\label{lum}
\end{equation}
is a luminosity which carries off the available rotational energy
$a_*Mc^2$ on the timescale $\tau$.

For a non--negligible Kerr spin parameter $a_* < 1$ the black hole's
rotational energy is comparable with its total rest--mass energy
$Mc^2$. Hence a significant misalignment of ${\bf J}$ and ${\bf B}$
offers an energy reservoir sufficient to power a gamma--ray
burst. However this energy is released only if the sources of the
magnetic field remain essentially fixed on the timescale
(\ref{4}). This depends on the relative importance of the hole and
source angular momenta $J, J_{\rm sources}$. If $J_{\rm sources} < J$,
the timescale (\ref{4}) and the corresponding energy release both
reduce by a factor $\sim J/J_{\rm sources}$ (King \& Lasota, 1977). Hence
significant extraction of black hole spin energy occurs only if
\begin{equation}
J_{\rm sources} > J
\label{5}
\end{equation}

\section{Gamma-ray Bursts}

Condition (\ref{5}) limits the applicability of the alignment
torque. Although Kim et al. (2003) note that the alignment timescale
(\ref{4}) is dimensionally the same as the BZ timescale, very little
of the black hole spin energy would emerge in a misaligned version of
the BZ picture before alignment was complete, because the sources of
the field (an accretion disc) have so little angular momentum compared
with the hole.

One obvious case where the sources of a strong external field have
high angular momentum and inertia occurs in a gamma--ray burst. In
the hypernova picture (Woosley 1993; McFadyen \& Woosley, 1999;
Paczy\'nski 1998) the degenerate core of an evolved and
rapidly--spinning star collapses. The central regions support
equipartition magnetic fields $\sim 10^{15}$~G within a few
gravitational radii of the spinning black hole forming in the
centre, and have high mass and angular momentum. Observations of
magnetars with fields of this order support this idea. The
well--attested occurrence of neutron--star kicks shows that core
collapse can be significantly anisotropic. Moreover the kick must
produce a spin (Spruit \& Phinney, 1998) uncorrelated with the
rotation of the stellar core. If the collapse forms a black hole
rather than a neutron star, one can expect cases where the hole's
spin is misaligned with the core rotation and thus presumably the
magnetic field. Unless this field varies over a lengthscale $\sim
R_g$ we can regard it as approximately uniform near the hole.
Under such conditions the alignment timescale (\ref{4}) is
\begin{equation}
\tau_{\rm GRB} = 4\times 10^3m^{-1}B_{15}^{-2}~{\rm s},
\label{6}
\end{equation}
where $m = M/\msun$ and $B_{15} = B/10^{15}~{\rm G}$. Even with weaker
fields $B_{15} \sim 0.1$ this is of order the rest--frame duration of
a gamma--ray burst with $\gamma \sim 100$ (cf Kim et al. (2003).

\section{The Prompt Radiation Pattern}

Press (1972) and King \& Lasota (1977) calculate the alignment torque
of a Kerr black hole, for stationary scalar and magnetic fields
respectively. In reality both the gravitational field and the
perturbing scalar or magnetic field must themselves vary on the
timescale (\ref{4}), implying the emission of both gravitational and
electromagnetic radiation. (This is obvious from the fact that both
fields change significantly between the start and end of the
alignment process.) The stationary approximation is still correct for
the purpose of calculating the torque; both fields
adjust on the light--travel time, and thus pass through a sequence of
stationary states. However this approach cannot give us the detailed
form of the radiation fields. Direct calculation of these fields is a
formidable undertaking, as testified by the extremely laborious nature
of the torque calculation in the stationary magnetic case (the author's
handwritten algebra for the 1977 paper with Lasota covers some 65
pages of large--format computer paper). However we can deduce the
nature of the electromagnetic emission by a simple argument.

The torque {\bf N} acts entirely on the misaligned component
$J_{\perp}$. The {\it net} torque exerted by the Poynting field
therefore consists of a couple acting about this axis. We can express
this as two oppositely--directed beams on each side of the axis, each
carrying radiation pressure force $L_p/c$ in a direction orthogonal to
$J_{\perp}$, separated by a lever arm. Since the torque ${\bf N}$
vanishes if the hole does not rotate, the lever--arm associated with
the $L/c$ beams must be of order the ergosphere radius, which is
itself of order the gravitational radius $R_g = GM/c^2$. The total
radiation field may also have a component $L_0$ with no net lever arm
(e.g. isotropic, or reflection--symmetric).

Since the $L_p$--beams exert the torque with lever--arm $R_g$ we have
from (\ref{ntau}) that
\begin{equation}
R_g{L_p\over c} \sim N \sim R_g{L\over c}
\label{7}
\end{equation}
i.e. the luminosity $L_p$ in the two opposing beams is of order the
total luminosity $L \sim a_*Mc^2/\tau$ released by the alignment
process. By symmetry the beams must be either parallel or orthogonal
to ${\bf B}$ as well as orthogonal to $J_{\perp}$. Alignment must
leave the far field unchanged and simply affect the pinching of the
field by the hole (cf King, Lasota \& Kundt, 1975), i.e. the component
orthogonal to the asymptotic field direction. Since the magnetic field
of an electromagnetic wave is transverse, this requires the beams to
be emitted parallel to this asymptotic field direction (and with a
definite polarization pattern).

This argument shows that a large fraction of the electromagnetic
luminosity released during magnetic alignment is emitted in an
opposed pair of narrow radiation beams (width $\sim$ ergosphere
radius $\sim R_g$) along the asymptotic magnetic field direction,
which in practice must have a slight intrinsic spread because of
the deviation from a completely uniform field. (The radiation does
not of course travel along field lines, but is simply emitted in
directions parallel to the far field.) The photon picture of this
emission is extremely simple: the misaligned angular momentum
component $J_{\perp}$ `spins off' photons from each side of the
ergosphere, parallel to the asymptotic field lines.

\section{Observational Appearance}

For a black hole mass $M \sim 10\msun$ forming in a degenerate core of
radius $\sim 10^9$~cm the prompt luminosity beams are directed over
solid angle $\sim 10^{-5}$. They presumably drive out core matter from
a pair of tubes with roughly these diameters, i.e. about $10^{-5}$ of
the core mass. An important point here is that the beams are
automatically directed along fieldlines in the region of strong field,
making it much easier to drive matter out. The alignment process thus
naturally satisfies the baryon--loading constraint (\ref{1}). Even if
the isotropic luminosity $L_0$ is of the same order as the beam
luminosity $L$, geometrical dilution means that it must produce much
lower Lorentz factors and thus be observationally insignificant.

\section{Discussion}

I have suggested that in a hypernova the black hole may form with its
spin misaligned with the strong magnetic field anchored in matter
further out in the core. Under these conditions magnetic alignment can
extract a significant fraction of the hole's rest--mass energy. I have
argued that much of this energy appears in two narrow,
oppositely--directed beams. These beams have properties making them
very suitable as the prompt emission for this type of gamma--ray
burst.

Similar conditions might well hold in a gamma--ray burst caused by the
coalescence of a black hole and a neutron star with a
magnetar--strength field. In both cases the misaligned angular
momentum component $J_{\perp}$ fixes the total energy reservoir. The
energy therefore comes from the kick process producing this
misalignment, and thus ultimately from the stellar spin and the
asymmetry of the core collapse.

One never observes the prompt emission (as opposed to the
fireball) from a gamma--ray burst, which makes observational tests
of this idea necessarily indirect. The considerable range of
observed neutron--star kick velocities does however suggest that
if alignment is the engine for gamma--ray bursts, these form at
least an one--parameter family.

Research in theoretical astrophysics at Leicester is supported by a
PPARC rolling grant. I thank Jim Pringle and Melvyn Davies for helpful
discussions. I gratefully acknowledge a Royal Society Wolfson Research
Merit Award.


\end{document}